\documentclass{article}
\usepackage[english]{babel}
\usepackage[margin=1in]{geometry}
\usepackage{multicol}
\usepackage{background}

\SetBgScale{4}
\SetBgColor{gray}
\SetBgAngle{90}
\SetBgContents{arXiv: some other text goes here}
\SetBgPosition{-1.5,-10}
\usepackage[square,numbers]{natbib}
\usepackage{longtable}
\usepackage{amsmath}
\usepackage{amsfonts}
\usepackage{amssymb}
\usepackage{makeidx}
\usepackage{graphicx,rotating,booktabs}
\usepackage{subfig}
\usepackage{color}
\usepackage{lmodern}
\usepackage{kpfonts}
\usepackage{xfrac}
\usepackage{array}
\usepackage{url}

\newcolumntype{L}[1]{>{\raggedright\let\newline\\\arraybackslash\hspace{0pt}}m{#1}}
\newcolumntype{C}[1]{>{\centering\let\newline\\\arraybackslash\hspace{0pt}}m{#1}}
\newcolumntype{R}[1]{>{\raggedleft\let\newline\\\arraybackslash\hspace{0pt}}m{#1}}
\usepackage{longtable}
\usepackage[verbose]{placeins}

\usepackage[utf8]{inputenc}

\usepackage{setspace,caption}
\usepackage{lipsum}
\usepackage{mwe}    
\usepackage{subfig}
\usepackage{float}
\usepackage[linesnumbered, ruled]{algorithm2e}
\SetKwRepeat{Do}{do}{while}%
\usepackage{algpseudocode}
\usepackage{comment}
\IncMargin{1em}

\usepackage{xr}
\begin{document}
 
{\centering
 
{\bfseries\Large Trans-Allelic Model for Prediction of Peptide:MHC-II Interactions \bigskip}

A. M. Degoot\textsuperscript{1,2,3} , Faraimunashe Chirove\textsuperscript{2} , and Wilfred Ndifon\textsuperscript{1} \\
{\itshape
	\textsuperscript{1} African Institute of Mathematical Sciences-(AIMS), 6 Melrose, Muizenberg,  Cape Town, South Africa. \\
	\textsuperscript{2}School of Mathematics, Statistics and Computer Science,
	University of KwaZulu-Natal, Scottsville ave, Pietermaritzburg, South Africa \\
	\textsuperscript{3}DST-NRF Centre of Excellence in Mathematical and Statistical Sciences (CoE-MaSS), Wits 2050,  Gauteng,  South Africa \\
	\normalfont \today 
 
   }
}

\begin{abstract}
 Major histocompatibility complex class two {(MHC-II)} molecules are trans-membrane proteins and key components of the cellular immune system. 
Upon recognition of foreign peptides  expressed on the MHC-II binding groove,  helper T cells  mount an immune response against   invading pathogens. Therefore, mechanistic identification and knowledge of  physico-chemical features that govern  interactions between  peptides and MHC-II molecules is useful for the  design of effective  epitope-based vaccines, as well as for  understanding of immune responses.
In this paper, we present a comprehensive trans-allelic prediction model, a generalized version of our previous biophysical model, that can predict peptide interactions for all  three human MHC-II loci (HLA-DR, HLA-DP and HLA-DQ), using both peptide sequence data and structural information of MHC-II molecules. The  advantage of this approach over  other machine learning models is  that it offers a simple and plausible physical explanation for peptide-MHC-II interactions.  
 We train the model using a benchmark experimental dataset, and measure its predictive performance using novel data. Despite its relative simplicity, we find that the model has comparable performance to the state-of-the-art method. Focusing on the physical bases of peptide-MHC binding, we find support for  previous theoretical predictions about the contributions of certain binding pockets to the binding energy. Additionally, we find that binding pockets $P4$ and $P5$ of HLA-DP, which were not previously considered as primary anchors, do make strong contributions to the binding energy. Together, the results indicate that our model can serve as a useful complement to alternative approaches to predicting peptide-MHC interactions.
\end{abstract}

\section{Introduction}
Major histocompatibility complex class two {(MHC-II)} molecules are surface proteins that exist on the  membrane of  antigen presenting cells-{(APCs)} such as macrophages, dendritic cells and B cells. They bind  short peptide fragments derived from exogenous proteins and present them to $CD4^{+}$ helper T cells.  Upon the recognition of foreign peptides among those presented by MHC-II molecules, the helper-T cells will initiate  proper adaptive immune responses, including enabling sufficient maturation of  B cells and cytotoxic $CD8^{+}$ T cells \cite{jana}. Therefore, the binding of peptide to MHC-II molecules is considered to be a fundamental and pre-requisite step in the initiation of adaptive immunity \cite{multirta, book1}. As such,  mechanistic identification of the basic determinants of peptide-MHC-II interactions presents potential for understanding the immune system's mechanisms and improving the process of designing peptide- and proteins-based vaccines.

MHC genes for humans, referred to as Human Leukocyte Antigen-({HLA}), are among the most  polymorphic genetic elements found within a long continuous stretch of DNA on chromosome 6 \cite{kuby}. Such high polymorphism reflects the immense contribution of MHC molecules to the adaptive immune system and underpins  their capacity to recognize a wide range of pathogens. Nonetheless, some  viruses, such as hepatitis C, avian/swine influenza and  human immunodeficiency virus (HIV), undergo extensive mutations that allow them to   escape recognition by the MHC molecules \cite{evac}.  MHC genes can be  divided into HLA class I, II and III. Molecules corresponding to HLA class I  are A, B and C;  HLA class II molecules  are DP, DQ and DR; HLA class III genes encode for  several other immune related proteins and provide support for the former two classes  \cite{jana, kuby}.

MHC-II molecules  account for  the likelihood of organ transplantation and there are well-established associations between many disorders and particular classes of MHC-II molecules. These  include the contribution of HLA-DQ genes to insulin-dependent diabetes \cite{dbt}; HLA-DR  genes to multiple-sclerosis; and narcolepsy \cite{mt} along with other autoimmune diseases  resulting from degeneracy and misregulation in the process of peptide presentation \cite{AID}. Moreover,  genetic and epidemiological data have implicated MHC-II molecules  in  susceptibility to many infectious diseases such as HIV/AIDS,  malaria \cite{hiv} and cancer
\cite{cancer}.

Experimental assays for prediction of peptide-MHC-II interactions are often  faced with  important obstacles, including substantial resources needed for laboratory work,  high  time and labour demands. This is  the case in particular, for experimental work aimed at finding out which promiscuous epitopes  bind to specific MHC molecules,  a necessary step in the design  of peptide-based vaccines which protect against a broad range of pathogen variants.   Computational methods, which are more efficient and less costly than  biological assays,  have been employed to complement these  assays.   Due to advances in sequencing technologies, immunological data has grown at an unprecedented pace and  continues to accrue.  This has been exploited in  systematic  computational analyses of  genomes of multiple pathogens  to determine the immunoprotective parts that can induce a potent immune response. The results have been  the design and development of  new vaccine candidates against HIV,  influenza and  other hyper-variable viruses \cite{apps}.  Use of computational methods has significantly reduced experimental effort and costs by up to $85\%$ \cite{ivan}.

Many immunoinformatics methods for prediction of peptide-MHC interactions, for both class I and II, have been developed based on  machine learning approaches such as simple pattern motif \cite{motif}, support vector machine (SVM) \cite{svm}, hidden Markov model (HMM) \cite{hmm} , neural network (NN) models \cite{pan, pan3}, quantitative structure–activity relationship (QSAR) analysis \cite{qsar}, structure-based methods, and  biophysical methods \cite{multirta, rta, ours, siadt}.  These methods can be divided into two categories, namely, intra-allele (allele-specific)  and trans-allele (pan-specific) methods. Intra-allelic methods are trained  for a specific MHC allele variant on a limited set of experimental peptide binding data and applied for prediction of peptides binding to that allele. Because of the extreme polymorphism of MHC molecules, the existence of thousands of allele variants, especially for HLA-II genes, combined with the lack  of sufficient experimental binding data, it is impossible to build a prediction model for each allele. Thus,  trans-allele and general purpose \cite{panr1}  methods  like \textit{MULTIRTA} \cite{multirta}, \textit{NetMHCIIpan} \cite{pan3} and \textit{TEPITOPEpan} \cite{tepitopepan}   have been developed using  richer peptide binding data expanding over many alleles or even   across species \cite{pan3}. The trans-allelic models are often designed to extrapolate either structural similarities or  shared   physico-chemical binding determinants  among HLA genes, in order to predict  affinities for alleles  that are not even part of the training dataset. These models generally have better predictive performances and a wide range of potential applications  compared to    the   intra-allelic models. 

Most of the  existing trans-allelic models for MHC-II  are  extended versions of their earlier intra-allelic counterparts: TEPITOPEpan \cite{tepitopepan} was extended from \textit{TEPITOPE} \cite{tepitope}; \textit{MULTIRTA} \cite{multirta}  evolved from \textit{RTA} \cite{rta}; and the series of \textit{NetMHCIIpans} (1.0,  2.0, 3.0, and 3.1) \cite{pan, pan2, pan3, pan3.1} were generalized from the  NN-align \cite{smmalign} method.  In the same vein, in this paper we present a trans-allele method, an extension of our previous method \cite{ours}, for prediction of peptide-HLA class II interactions based on biophysical ideas. 

The remarkable strength of the method presented here over other existing advanced data driven approaches is it's physical basis. We formulate the process of binding affinity between peptide and MHC-II molecule as an inverse problem of statistical physics. From the observable macroscopic  (bound and unbound ) states of experimental data we compute the microscopic parameters ( Hamiltonians for amino acid residues  involved in the interaction) that govern the system. In fact,   many  problems in computational biology  can be solved in a such way \cite{moh, inverse}, taking advantage of  the availability of vast amount of genomic data and high resolution structural information. Solutions obtained using this approach are more plausible and physically interpretable than those obtained using mere sequence-based methods \cite{ multirta, ours}. Additionally, because  sparsity is a hallmark feature of biological processes, we adjust the model's parameters via incorporating an $\mathsf{L_1}$ regularization term into the model. The $ \mathsf{L_1}$ constraint, commonly named \textit{Lasso}, promotes sparsity and improves the predictive performance of the model on  novel data.

The rest of this paper is organized as follows: In Section \ref{sec2}, we describe the idea of MHC-II polymorphic residue groups, which is employed to capture  structure similarity among MHC-II alleles. In Section \ref{sec3}, we define our methodology and formulate the learning function. After that we briefly describe  the benchmark dataset used to test the predictive performance of the model in Section \ref{data} and present the results in Section \ref{sec4}. Finally, in Section \ref{sec5} we summarize and discuss our results and compare our method  with the state-of-the art method. 

\section{MHC-II Polymorphic Residue  Groups}
\label{sec2}
Crystal structures revealed that  an MHC molecule  is a combination of two domains, an $\alpha$ helix and a $\beta$ sheet, linked together  to form a Y-shaped groove which is used to locate peptides, and both domains equally contribute to the binding affinity. For HLA-I molecules, the $\beta$ domain is largely conserved and variation occurs mostly in the $\alpha$ domain. On the other hand, polymorphism occurs in both domains of HLA-II molecules; except for HLA-DR alleles, where the variation takes place in  the $\beta$ domain.  Additionally, the peptide binding groove of the HLA-II is open at both ends, which allows  binding peptides of variable lengths, ranging from 9 to 30 amino acid residues, or even an entire protein \cite{pan3.1, fact1}. This is in contrast to the peptide binding groove of the HLA-I alleles, which accommodate only short peptides of lengths ranging from 8 to 11 amino acids. This flexible constraint on peptide lengths together with it's immense polymorphism, contribute to a lower predictive performance of computational methods for  peptide-MHC-II interactions   compared to MHC-I methods \cite{multirta, panr1}. 

The  notion of MHC polymorphic residue groups, introduced by  Bordner et. al \cite{multirta}, is based on a simple observation of an intrinsic (independent of peptide) feature of the MHC-II binding groove. Although a peptide could bind to an MHC-II molecule in various registers, due to the open-ended nature of the MHC-II binding groove, the strength of the binding affinity is primarily determined by 9 residues occupying the binding groove pockets. Interestingly,  most of polymorphism in MHC-II genes occurs at these binding pockets (see the discussion in Section \ref{sec5}). 

From the limited available crystallographic structural data of peptide-MHC-II complexes for a few MHC-II molecules from the Protein Data Bank-(\text{PDB})  \cite{pdb} (summarized in \textbf{Table \ref{S-tab}} in the supplementary material),  sets of  important  positions for the polymorphic residues in the  binding groove  that  contact one or more   peptide binding cores and are within a distance of not more than 4\AA \hspace{0.1cm}   \cite{multirta, pan3, hundi}   in one or more of the MHC-II complex structures can be  extracted. Then, by extrapolating the similarities among MHC molecules, their corresponding residues in different genes are determined using multiple sequence analysis, \text{(MSA)} \cite{msa}. Exploiting the fact that HLA-DR alleles are polymorphic only in the $\beta$ domain and have the same $\alpha$ domain, the polymorphic residue groups for HLA-DR are extracted from its $\beta$ domain sequences.  Similarly, relying on the assumption of symmetric contribution between $\alpha$ and $\beta$ domains  to the binding affinity \cite{multirta}, residue groups for HLA-DP and HLA-DQ  were also extracted from the $\beta$ domain.

Next, the set of polymorphic residues that always co-occur at the specified positions are clustered into the same group. The rationale of clustering polymorphic residue groups, rather than individual residues, is to avoid over-parametrization of the model. \textbf{Table \ref{S-tab1}} in the supplementary material  shows such polymorphic residue groups for  HLA-DRB, HLA-DP and  HLA-DQ alleles, assembled by the procedures described above.
\section{Trans-Allele Model}
\label{sec3}
In our previous intra-allele model \cite{ours} the probability of peptide  $\mathsf{P^{(k)}}$ to bind an MHC molecule $\mathsf{M^{(T(k))}}$ was computed as follows:
\begin{equation}
\label{e1}
\pi (\mathsf{P^{(k)}, M^{(T(k))}}) = \frac{1}{ 1+ e^{\mathsf{\delta E^{(k)}}}},
\end{equation} 
where $\mathsf{\delta E^{(k)}}$ is the change in binding energy in terms of  the sum of  the differences of first- and second-order Hamiltonians between the bound  and unbound states. Specifically, $\mathsf{\delta E^{(k)}}$ is given by
\begin{eqnarray}
\label{e2}
\mathsf{\delta E^{(k)}} = & \underbrace{\sum\limits_{i=1}^{|\mathsf{P^{(k)}}|} \delta H^{(1)}(\mathsf{a}_i) + \sum\limits_{i=1}^{9} \delta H^{(1)}(\mathsf{b}_i) +  \delta S }_\text{first-order Hamiltonians}  +   \overbrace{ \underbrace{ \sum\limits_{i=1}^{|\mathsf{P^{(k)}}|} \sum\limits_{j=1}^{9} \sum\limits_{r=1}^{\mathsf{R}}\delta H^{(2)}(\mathsf{a}_{ir}^{(k)},\mathsf{b}_j)}_\text{second-order Hamiltonians} }^\text{per residue-residue interactions},  
\end{eqnarray}
in which $|\mathsf{P^{(k)}}|$ is the length of peptide $\mathsf{k}$, $\mathsf{R}$ is the number of all possible configurations (registers) in which the peptide  binds to the particular MHC molecule, and $\mathsf{ \delta S}$ is the difference in entropy between the bound  and unbound states. 

For the trans-allele model, two changes were introduced into the second term of  \textbf{Eq}(\ref{e2}). First, instead of residue-residue interaction, $\mathsf{\delta H^{(2)}(a_{ir}^{(k)}, b_j)} $,   with $\mathsf{a_{ir}^{(k)}}$ on the peptide sequence  and $\mathsf{b_j}$ on the MHC binding pocket, we rather  focus on residue-polymorphic group interaction, $\mathsf{\delta H^{(2)}(\mathsf{a}_{ir}^{(k)},\mathsf{g}_{jn})}$, where $\mathsf{g_{jn}}$ is residue group number $n$ of position $j$ as defined in Section \ref{sec2}.
Next, we introduce a binary operator $\mathsf{T(k, j, n)}$ that  equals  $1$ if the MHC molecule type,  $\mathsf{M^{(T(k))}}$, corresponding to peptide  $\mathsf{P^{(k)}}$ contains polymorphic residue group $n$ at the set of pre-determined positions  of pocket $j$, and  equals $0$ otherwise. Hence, $\mathsf{\delta E^{(k)}}$ is given by
\begin{eqnarray}
\label{e3}
\mathsf{\delta E^{(k)}}=& \underbrace{\sum\limits_{i=1}^{|\mathsf{P^{(k)}}|} \delta H^{(1)}(\mathsf{a}_i) + \sum\limits_{i=1}^{9} \delta H^{(1)}(\mathsf{b}_i) +  \delta s }_\text{first-order Hamiltonians}  +  \overbrace{ \underbrace{ \sum\limits_{i=1}^{|\mathsf{P^{(k)}}|} \sum\limits_{j=1}^{9} \sum\limits_{r=1}^{\mathsf{R}} \sum \limits_{n=1}^{\mathsf{G(j)}}\delta H^{(2)}(\mathsf{a}_{ir}^{(k)},\mathsf{g}_{jn}) \mathsf{T(k, j, n)}}_\text{second-order Hamiltonians} }^\text{per residue-group interactions},  
\end{eqnarray}
where $\mathsf{G(j)}$ is the number of polymorphic residue groups for  binding pocket $j$. Column two of \textbf{Table \ref{S-tab1}} in the supplementary material shows $\mathsf{G(j)}, j = 1,2, \ldots 9$, for HLA-DR, HLA-DP, and HLA-DQ alleles.

Let $\Delta$ denote the model's parameters. Using Equations (\ref{e1}) and (\ref{e3})  we formulate, through the maximum likelihood approach,  the following cost function:

\begin{equation}
\label{e4} 
\mathcal{L}(\mathsf{P}, \mathsf{M}| \mathsf{\Delta}) = \underset{\{ \mathsf{\Delta}\}}{\mathrm{argmin}}\left(\sum\limits_{\mathsf{k}= 1}^{\mathsf{K}} \mathsf{G}^\mathsf{k}(\mathsf{\Delta^k}) + \mathsf{\lambda \mathcal{P}(  \mathsf{\Delta})}\right),
\end{equation}
where
$\mathsf{G}^\mathsf{k}(\Delta) $ is the empirical loss function given by 
\begin{equation}
\label{e5}
\mathsf{G}^\mathsf{k}(\Delta) = \mathsf{y}^\mathsf{k}\log(\pi^\mathsf{k}(\Delta)) + (1 -\mathsf{y}^\mathsf{k})\log(1 - \pi^\mathsf{k}(\Delta)),
\end{equation}
and $\mathsf{y}^\mathsf{k} \in \{0,1\}$ is the experimental value;  $y = 1$ for binding peptides and $y = 0$ for non-binding ones.
$\lambda \mathcal{P} (\Delta)$ is a regularization term with the form 
\begin{equation}
\label{e6}
\lambda \mathcal{P} (\mathsf{\Delta}) =\lambda  \left\vert\left\vert \mathsf{\Delta} \right\vert\right\vert_1  = \lambda\sum\limits_{i=1}^{d} \left\vert \Delta \right\vert, 
\end{equation}
where  $\mathsf{\lambda} > 0$ is a hyper-parameter and $d$ is the dimension of parameter vector $\Delta$, which  varies  depending on the type of MHC-II molecule. The $L_1$ constraint penalty term $\mathcal{P} (\mathsf{\Delta})$, also known as Lasso \cite{lasso}, has an important role in the model. As the model is defined on a large number of parameters ( $d = 2321$ , $ 561$ and $ 401$ for HLA-DR, HLA-DP and DQ molecules, respectively) a few parameters are expected to contribute to the binding affinity while the  rest are expected to be noisy. Lasso has the capability to filter out the noisy parameters by inducing sparsity in the model, as it shrinks most of the parameter values to zero,  and avoids data over-fitting.  The hyper-parameter $\lambda$ controls the degree of  sparsity of the model; the larger the value of $\lambda$ the more sparse the model.
\textbf{Eq}(\ref{e4}) is a non-linear and non-smooth function; due to the  $L_{1}$ constraint. But it is a convex function  and we solved it, after quadratic approximation, by means of an iterative, cyclic coordinate descent approach using a soft-thresholding operator. This learning function takes the form of a generalized linear model and the algorithm we used to solve it is both fast and efficient. Details of this optimization method are found in Friedman et. al \cite{glm} and are summarized in the supplementary material.

\section{Binding Affinity Dataset}\label{data}

The model has been developed by using both quantitative peptide binding data and MHC-II molecule sequences. We obtained a total of $51023$ peptide-binding data for $24$ HLA-DR, $5$ HLA-DP and $6$ HLA-DQ   from the IEDB database \cite{iedb}, which is, to the best of our knowledge, the largest benchmark dataset publicly available in this field. This dataset was used to develop NetMHCIIpan \cite{pan3}, the state-of-the-art method. The binding affinities  data  were given  in the form of log-transformed  measurements of the IC$_{50}$ (half maximum inhibition concentration) according to the formula   $1- \log(IC_{50})/\log(50,000)$ \cite{log}. We dichotomized these data using a moderate threshold of IC$_{50}$ $500$ nM ( $\equiv 0.426$ of log-transformed data).  Peptides with  $IC_{50}$  less than or equal $500$ nM  ($\geq 0.426$ of log-transformed value ) were considered as binders, and non-binders otherwise. 

Amino acid sequences for the MHC-II alleles used in this study were obtained from the \textbf{EMBL-EBI} online-database \cite{ebi},  (\url{ftp://ftp.ebi.ac.uk/pub/databases/ipd/imgt/hla/fasta} ). \textbf{Table \ref{tab2}} gives a summary of the peptide binding dataset used to develop the method.
\begin{center}
	\begin{longtable}{L{5cm}  C{2.3cm} C{2.3cm}  C{2.3cm}  C{2.5cm}}
		\caption{ Overview of the MHC-II peptide binding data  utilized in this study.}  \label{tab2}\\
		\toprule[1mm]
		\textbf{Allele Name} & \textbf{HLA-Index}& $\#$\textbf{of peptides} & $\#$\textbf{of binders} & $\%$ \textbf{of binders}\\
		\endfirsthead
		\midrule[0.8mm]
		\endhead
		
		\multicolumn{5}{c}%
		{{\bfseries \tablename\ \thetable{} -- continued from previous page}} \\
		\midrule[0.4mm] 
		\endhead
		
		\multicolumn{5}{c}{{Continued on next page}} \\
		\endfoot
		\endlastfoot
		\midrule[1.5mm]
		\multicolumn{5}{c}{\textbf{HLA-DR Molecules}}  \\ \midrule[1mm]
		$DRB1^*01:01$ & HLA00664 &$7685$ & $4382$ &  $57.02$\\  \midrule
		$DRB1^*03:01$ & HLA00671 &$2505$ & $649 $ &  $25.91$\\ \midrule
		$DRB1^*03:02$ & HLA00673 &$148$  & $44  $ &  $29.73$\\ \midrule
		$DRB1^*04:01$ & HLA00685 &$3116$ & $1039$ &  $33.31$\\ \midrule
		$DRB1^*04:04$ & HLA00689 &$577$  & $336 $ &  $58.23$\\ \midrule
		$DRB1^*04:05$ & HLA00690 &$1582$ & $627 $ &  $39.63$\\ \midrule
		$DRB1^*07:01$ & HLA00719 &$1745$ & $849 $ &  $48.65$\\ \midrule
		$DRB1^*08:02$ & HLA00724 &$1520$ & $431 $ &  $28.36$\\ \midrule
		$DRB1^*08:06$ & HLA00732 &$118$  & $91  $ &  $77.12$\\ \midrule
		$DRB1^*08:13$ & HLA00739 &$1370$ & $455 $ &  $33.21$\\ \midrule
		$DRB1^*08:19$ & HLA00745 &$116$  & $54  $ &  $46.55$\\ \midrule
		$DRB1^*09:01$ & HLA00749 &$1520$ & $621 $ &  $40.86$\\ \midrule
		$DRB1^*11:01$ & HLA00751 &$1794$ & $778 $ &  $43.37$\\ \midrule
		$DRB1^*12:01$ & HLA00789 &$117$  & $81  $ &  $69.23$\\ \midrule
		$DRB1^*12:02$ & HLA00790 &$117$  & $79  $ &  $67.52$\\ \midrule
		$DRB1^*13:02$ & HLA00798 &$1580$ & $493 $ &  $31.20$\\ \midrule
		$DRB1^*14:02$ & HLA00834 &$118$  & $78  $ &  $66.20$\\ \midrule
		$DRB1^*14:04$ & HLA00836 &$30 $  & $16  $ &  $53.33$\\ \midrule
		$DRB1^*14:12$ & HLA00844 &$116$  & $63  $ &  $54.31$\\ \midrule
		$DRB1^*15:01$ & HLA00865 &$1769$ & $709 $ &  $40.08$\\ \midrule
		$DRB3^*01:01$ & HLA00887 &$1501$ & $281 $ &  $18.72$\\ \midrule
		$DRB3^*03:01$ & HLA00902 &$160$  & $70  $ &  $43.75$\\ \midrule
		$DRB4^*01:01$ & HLA00905 &$1521$ & $485 $ &  $31.89$\\ \midrule
		$DRB5^*01:01$ & HLA00915 &$3106$ & $1280$ &  $41.21$\\ \midrule[0.8mm]
		\multicolumn{5}{c}{\textbf{HLA-DP Molecules}}  \\ \midrule[0.8mm]
		$DPA1^*01:03-DPB1^*02:01$ & HLA00517 &  $1404$ & $538 $ &  $38.32$\\ \midrule
		$DPA1^*01:03-DPB1^*04:01$ & HLA00521 &  $1337$ & $471 $ &  $35.23$\\ \midrule
		$DPA1^*02:01-DPB1^*01:01$ & HLA00514 &  $1399$ & $597 $ &  $42.67$\\ \midrule
		$DPA1^*02:01-DPB1^*05:01$ & HLA00523 & $1410$ & $443 $ &  $31.42$\\ \midrule
		$DPA1^*03:01-DPB1^*04:02$ & HLA00522 & $1407$ & $523 $ &  $37.17$\\ \midrule[0.8mm]
		\multicolumn{5}{c}{\textbf{HLA-DQ Molecules}}  \\ \midrule[0.8mm]
		$DQA1^*01:01-DQB1^*05:01$ & HLA00638 &$1739$ & $522 $ &  $30.02$\\ \midrule
		$DQA1^*01:02-DQB1^*06:02$ & HLA00646 &$1629$ & $813 $ &  $49.91$\\ \midrule
		$DQA1^*03:01-DQB1^*03:02$ & HLA00627 & $1719$ & $386 $ &  $22.46$\\ \midrule
		$DQA1^*04:01-DQB1^*04:02$ & HLA00637 & $1701$ & $559 $ &  $32.86$\\ \midrule
		$DQA1^*05:01-DQB1^*02:01$ & HLA00622 & $1658$ & $549 $ &  $33.11$\\ \midrule
		$DQA1^*05:01-DQB1^*03:01$ & HLA00625 & $1689$ & $863 $ &  $51.10$\\ \midrule[1mm]
		\textbf{Total} &          &$51023$&    $20255$ &  ${39.70}$\\
		\bottomrule[1.5mm]
		\caption{The first column gives the names of the $34$ genes used to develop the method, distributed as $24, 5, 6$ for HLA-DR,  HLA-DP and   HLA-DQ  genes respectively.  The second column represents the index for  each allele in the \textbf{EMBL-EBI} database \cite{ebi}. The third and fourth columns give the total number of peptide   and the number of binder peptides, receptively, per  allele.  The last column shows the percentage of binder peptides. Binder peptides were identified using an $IC_{50}$ binding cut-off of $500$ nM, as in previous studies \cite{multirta, pan, pan3, smmalign}. The last row presents the overall statistics for the last three columns.}
	\end{longtable}
	
\end{center}
\section{Results}
\label{sec4}
This section presents prediction results of the model obtained from the dataset of three MHC-II allotypes  as described in  Section \ref{data}. We applied a five-fold cross validation analysis to the model  and compared it against its intra-allelic version (\textbf{Table \ref{S-tabfold}} in the supplementary material). We also  examine its predictive performance on  data which were previously unseen by the model.

\subsection{Performance of the trans-allele model}
\begin{figure}[!h]
	\textbf{AUC ROC curves for the three loci}
	\centering	\includegraphics[scale = 0.85]{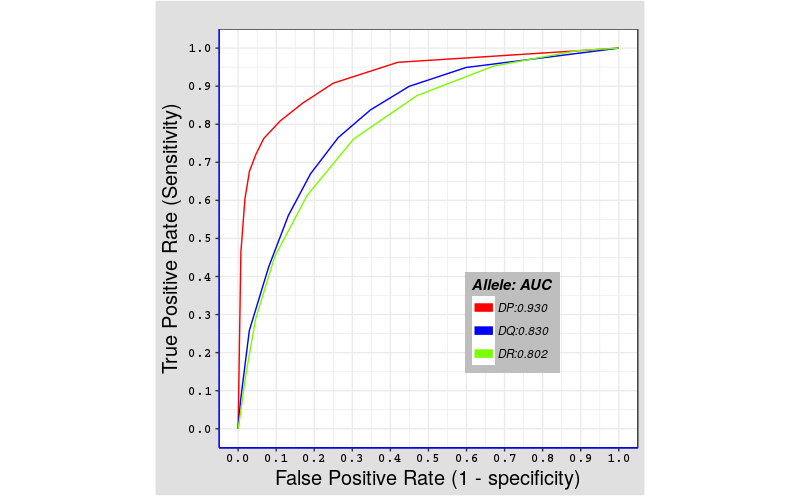}
	\caption{\textbf{(colours online)} Shows Five-fold cross validation results of the model using the benchmark dataset described in the Section \ref{data}.  Three ROC curves representing the three MHC-II loci covered in this study. The red curve for HLA-DP  with AUC value $= 0.930$, the blue curve for HLA-DQ  with AUC value $= 0.830$ and  the green curve for HLA-DR with AUC $= 0.802$.}
	\label{fig1}
\end{figure}

We tested the predictive performance of the model by using five fold cross validation.  Figure \ref{fig1} shows  results of the test done using alleles belonging to the three MHC-II loci considered in this study. The performance was measured in terms of area under the curve ({AUC})  \cite{auc} values,  which  range between $0$ and $1$. The higher the AUC value the better the predictive performance of model. Values below $0.5$ reflect a worse performance than a random test. The model has an excellent performance for HLA-DP alleles  (average  AUC value $= 0.930$),  and a good predictive power for both  HLA-DQ and HLA-DR alleles   (average AUC  values $= 0.830$ and $0.802$, respectively).

\subsection{Comparing the intra-allele vs trans-allele methods}
\begin{figure}[!h]
	\textbf{intra-allele vs trans-allele method}
	\centering	\includegraphics[scale = 0.81]{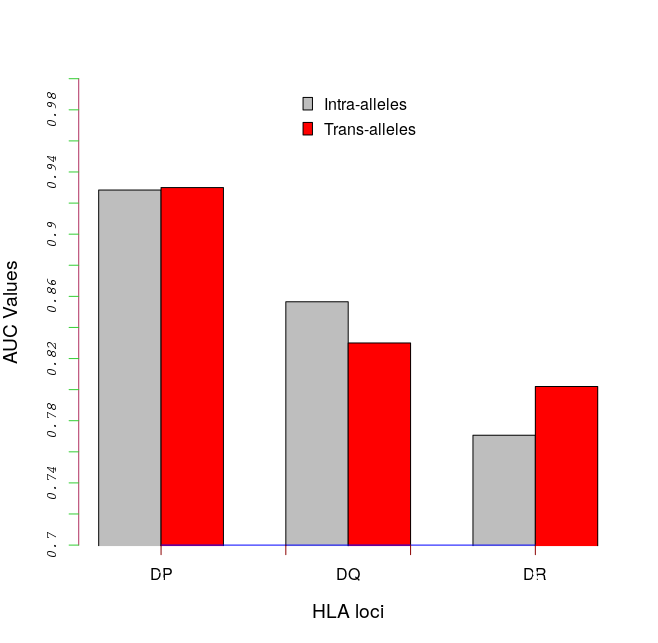}
	\caption{\textbf{(colours online)} Comparing results between the  intra-alleles    (gray bars) and the  trans-alleles (red bars)  methods  in terms of AUC values. These bars show that there is a significant increase in performance of the trans-allele method for HLA-DR molecules and decrease for HLA-DQ molecules compared to the intra-allele method. The difference in the HLA-DP loci is limited.}
	\label{fig2}
\end{figure}

\textbf{Table \ref{S-tabfold}} in the supplementary material shows AUC values obtained with the  intra-allele and  trans-allele versions of the model. For  the intra-alleles version, the model was evaluated on peptide binding data corresponding to an individual allele only. On average, the  performance of the trans-allele model is comparable to that of the intra-allele model for HLA-DP ($O.930$ vs $0.928$), it is worse for HLA-DQ ($0.830$ vs $0.857$) and it is better for HLA-DR ($0.780$ vs $0.771$) (Figure \ref{fig2}).

These results demonstrate two  important observations.   First, there is  a common binding preference among MHC-II loci,  which is the basis of all trans-allelic models, and that  has been successfully captured by the definition of MHC-II polymorphic groups  for   HLA-DP loci, and to a lesser extent for HLA-DQ and HLA-DR. Second,   the trans-allelic model is able to extrapolate  similarities among the MHC-II allotypes and achieve  good predictive performance.  As a result, the overall performance of the trans-allelic model is comparable to that of intra-allele model, even though the former model is applied on a much diverse set of MHC-II sequences. 

A decreased performance of the trans-allelic model  when compared with the intra-allelic method for HLA-DQ molecules is consistent with results reported in NetMHCIIpan \cite{pan3}.  Here we suggest that this is probably because of the limited structural information available for HLA-DQ alleles. In fact, because of this limited structural information there are  only $17$ polymorphic residue groups for all the $9$ binding pockets defined for HLA-DQ alleles. In contrast, there are  $25$ and $115$  polymorphic residue groups defined for HLA-DP and HLA-DR molecules, respectively. 

Another reason for the reduction of the trans-allelic model's performance for HLA-DQ alleles is that there is a large  sequence diversity of MHC-II molecules belonging to this locus.  We will examine the empirical support for this assertion in Section\ref{nn}.

\subsection{Prediction on a novel dataset}\label{nn}
We examined the predictive power  of the model on a blind dataset- i.e, a  dataset which was not used  in the training phase. More precisely, to make peptide binding predictions for a particular allele, we train the model on an entirely different allele. The allele used for training was chosen based on its similarity to the focal allele as quantified using  three different metrics: nearest neighbour, Hamming distance, and Leave-One-Out (LOO) approach. 

In the nearest neighbour approach the distance between two MHC molecules is defined \cite{pan} as  follows:

\begin{equation}
\label{ed1}
d(A,B) = 1 - \frac{S(A,B)}{\sqrt{S(A,A)S(B,B)}}
\end{equation}
in which $S(A,B)$ is the score of the {BLOSUM50} \cite{blosum}    metric  between amino acid sequences of $A$ and $B$. The {BLOSUM50} metric measures genetic distance between two sequences by quantifying the likelihood that one amino acid will be substituted	by another amino acid on   evolutionary time  scales. Hamming distance simply counts the different occurrences of corresponding amino acid residues between two sequences.
In both nearest neighbour and Hamming metrics, we train the model on peptide data belonging to the corresponding nearest allele to  parameterize  the model, then we assess its accuracy in terms of AUC values calculated based on peptide data belonging to the focal allele  using those parameters.

However, unlike the  TEPITOPE  and  the series of NetMHCIIpan  methods which defined nearest neighbour at pocket level,  we derive  both the nearest neighbour metric and the Hamming distance at residue level. Our choice is based on the fact that accounting for the entire MHC-II sequence provides a broader allele coverage \cite{multirta} and hence extend the model's applicability. Computing sequence similarity at residue level is  an intuitive and natural approach to perform comparative analysis of sequences rather than other artificial ways that may be more computationally efficient.   We found that  $71\% $ (for HLA-DR ),  $60\% $ ( HLA-DP ) , and   $ 67\% $ ( HLA-DQ )  of alleles used for training were consistent  between the residue-level and pocket-level approaches.  These statistics indicate that, as mentioned before,  most of MHC-II polymorphism occur at the binding pockets. 

\begin{figure}[!h]
	\textbf{Comparison results of the three metrics.}
	\centering	\includegraphics[scale = 0.9]{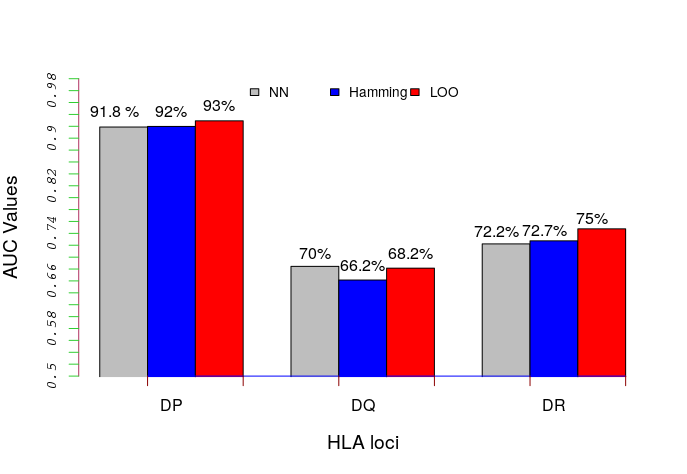}
	\caption{\textbf{(colours online)} Average performance results of the model in terms of AUC values for the three metrics: NN  approach  (gray bars), Hamming metric (blue bars) and the  LOO method (red bars). Except for HLA-DQ loci, the LOO approach significantly out performs the other two metrics. Such results indicate that this method performs better than a random test even for un-characterized MHC-II molecules.}
	\label{fig3}
\end{figure}

The  LOO approach involved partitioning data into two parts; the peptide binding data not belonging to the allele under consideration are used to learn the model's parameters and the remaining data, the peptide binding data belonging to the focal allele, are used as test data. Figure \ref{fig3} shows  a comparison of results from these three approaches (details are in \textbf{Table \ref{S-tab3}} in the supplementary material). The results show that, regardless of the metric we used, the trans-allele method has a high predictive power for HLA-DP allele and a moderate predictive power for the other alleles.

\begin{figure}[!h]
	
	\begin{tabular}{cc}
		\includegraphics[width=85mm]{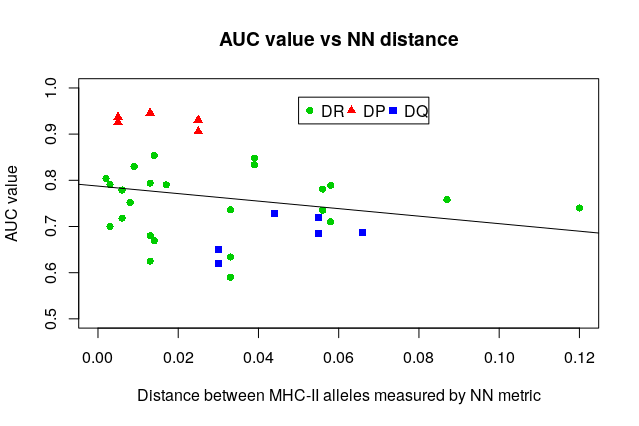} &  \includegraphics[width=85mm]{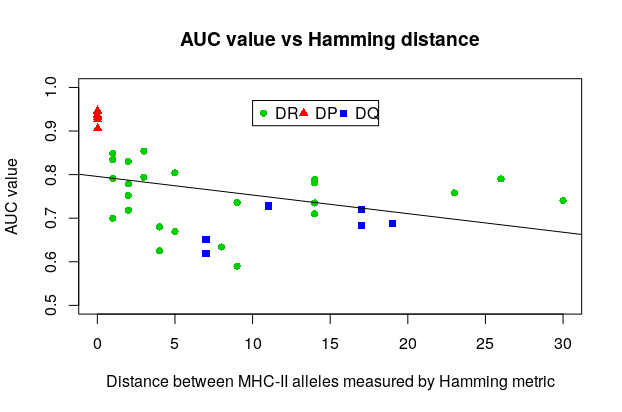} \\
		(a) NN metric & (b) Hamming metric
	\end{tabular}
	\caption{(colours online) Regression analysis of AUC values from the LOO approach as function of: \textbf{(a)} Nearest Neighbour  and \textbf{ (b)}  Hamming distances. Negative slope lines in both graphs obtained by the least square fit method, with p-values $0.185$ and $0.0.033$ for both metrics, respectively. These lines and p-values associated with were produced using glm2 package in R \cite{glm1}}
	\label{fig}
\end{figure}

The much higher predictive power for HLA-DP compared to the other alleles is likely due to the comparatively lower sequence diversity of HLA-DP alleles. To make this assertion more precise  we carried out a regression analysis by defining the AUC values from LOO approach as functions of both NN and Hamming metric distances. Figure \ref{fig} gives results of our analysis. As seen in Figure \ref{fig}, all HLA-DQ alleles fall below the least squares lines for both metrics (blue points). We also found that model performance  increases as the distance between alleles decreases, for example, see HLA-DP allele (red points). The authors of NetMHCIIpan  also arrived at  the same conclusion \cite{pan3}, but only for the NN metric.

\label{sec5}

\subsection{Analysis of the model's parameters}
In order to determine the key factors that contribute to the binding affinities for the  three MHC-II alleles considered in this study, we calculated the Hamiltonians  corresponding to each amino acid residue and the $9$ binding pockets of the MHC-II binding groove. These values were then averaged over the polymorphic residue groups defined for each pocket.

\begin{figure}[!h]
	
	\begin{tabular}{ccc}
		\includegraphics[width=60mm]{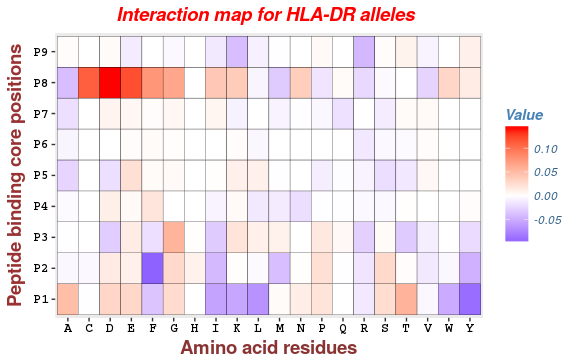} &   \includegraphics[width=60mm]{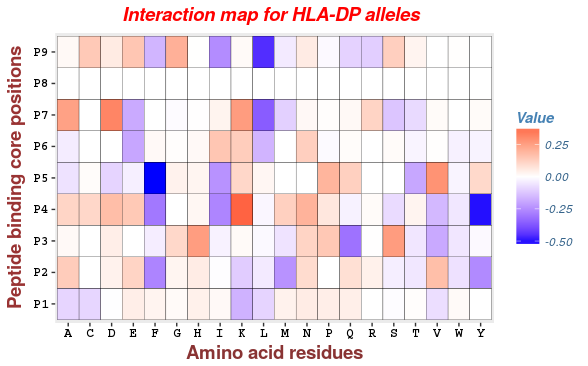} & \includegraphics[width=60mm]{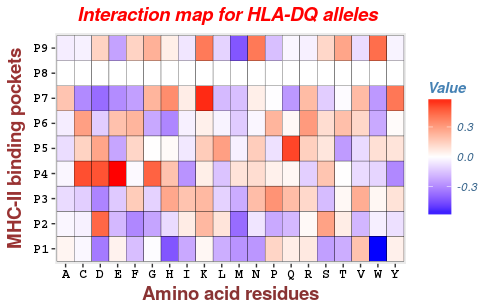} \\
		(a) HLA-DR & (b) HLA-DP & (c) HLA-DQ
	\end{tabular}
	\caption{(colours online) Interaction maps for: \textbf{(a)} HLA-DR,\textbf{ (b)}  HLA-DP and \textbf{(c)}  HLA-DQ molecules. The rows give the 9 anchor positions of MHC-II  binding groove and the columns give the peptide residues. The red entries marking for repulsive interactions (positive energy), whereas the blue entries marking for attractive interactions (negative energy). Note that most of the entries are zeros (white colour), an indication for the degree of the sparsity of the model.}
	\label{fig4}
\end{figure}

Analysis of HLA-DR parameters revealed that pocket $P1$ has  moderate attractive interactions with peptide (negative energies indicated by  blue colour in Figure \ref{fig4}), via  hydrophobic \textbf{(I, L, W, Y )} side chains and,  to lesser extent, via the aromatic \textbf{(F, W)} amino  acids  and a single   hydrophilic residue \textbf{(K)} . Remarkably, previous studies \cite{multirta, t3d}  arrived at a similar conclusion of a large tendency of position $P1$ toward interactions involving the hydrophobic side chains. The repulsive interactions (positive energies indicated by  red colour in Figure \ref{fig4}) of pocket $P1$ mostly occur with the hydrophilic side chains \textbf{(D, E, N, S, T)}  and the aliphatic residue \textbf{(A) }. Generally, most of the primary anchor pockets $(P1, P4, P6, P7, P9)$ confer attractive interactions, but  the pocket $P1$ makes the largest contribution. This is consistent with results obtained using the MULTIRTA method \cite{multirta}. Among the secondary anchors, we found that pocket $P2$ has attractive interactions with  aromatic \textbf{(F, Y)} and the hydrophobic \textbf{(I, M, Y)} side chains. The most repulsive interactions  come from the pocket $P8$,  which has a strong unfavourable interactions involving the side chains of residues \textbf{C, D, E, F, G, I, L, W,} and  \textbf{Y} ( see Figure \ref{fig4} \textbf{(a)}).

For HLA-DP, we found that  pocket $P9$  has significantly  attractive interactions involving the hydrophobic residue (\textbf{L}). This is consistent with the previous results of \cite{dpq}   (see Figure \ref{fig4} \textbf{(b)}). Also, we found that pockets $P4$ and $P5$ have important   attractive interactions with peptide via hydrophobic \textbf{(Y)} and  aromatic \textbf{(F)} side chains, respectively.  The  contributions of these two pockets were not reported in the study of Morten et. al  \cite{dpq},  which was specifically dedicated to HLA-DQ and HLA-DP alleles.  Furthermore, we found that the other two pockets $P1$ and $P6$, which were reported as primary anchors in that study, have a moderate  contribution to calculated bind energies ( see Figure \ref{fig4} \textbf{(b)}). 

The pattern of energetic contributions for HLA-DQ alleles is less ordered. There is no common pattern except the observation of significant attractive interaction of pocket $P1$ via the hydrophobic residue \textbf{(W)} and the repulsive interaction of pocket $P4$ via the side chains \textbf{C, E,} and \textbf{D} (see Figure \ref{fig4} \textbf{(c)}). This finding is in line with the observations of Morten et. al \cite{dpq}.

\subsection{Discussion}

Interactions between  peptides and MHC-II molecules are central to the adaptive immune system. Precise prediction and  knowledge of the physico-chemical determinants that govern such interaction is useful in designing  effective and affordable epitope-based vaccines, and in providing insights about the  immune system's mechanism as well as in understanding the pathogenesis of diseases.
In this study we have developed a trans-allelic model that can predict peptide interactions to the   three human MHC-II loci.  It  can be readily applied to MHC-II molecules of other species provided that relative structural information are available. This method is based on biophysical ideas, an alternative to the dominant machine learning approaches. 

The  model presented here is, in addition to NetMHCIIpan, only the second trans-allelic method that allows comprehensive prediction analysis of peptide binding to all three human MHC-II loci. Most  trans-allelic models for MHC-II peptides are restricted to HLA-DR and HLA-DP alleles. The TEPITOPEpan method \cite{tepitopepan}, which is  popular  among  immunologists and  is the  successor of a pioneer method in this field,  is limited to HLA-DR alleles.

In this work we employed the definition of MHC polymorphic residue groups of the MULTIRTA method \cite{multirta}, which is more intuitive and inclusive than the MHC pseudo sequences of NetMHCIIpan \cite{pan3}, in developing our trans-allelic model. Utilizing  new structural data for MHC-II complexes, which were not present when  MULTIRTA was being developed,  we extended that idea to cover all three human MHC-II loci.

\begin{figure}[!h]
	\centering	
	\textbf{Current model vs NetMHCIIpan}
	\includegraphics[scale = 0.9]{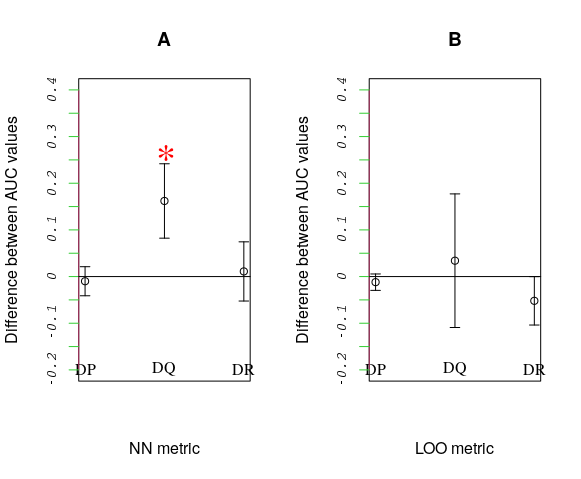}
	\caption{ Performance comparison between our model and NetMHCIIpan. Each model was used to predict the probability of peptide binding to query alleles belonging to each of three HLA loci (i.e. HLA-DP, HLA-DQ, HLA-DR) after training it using peptide-binding data for a different allele. The allele that was most similar to the query allele was used for training. As in previous work \cite{pan3}, similarity between HLA alleles was defined based on two metrics: nearest neighbor (NN) and leave-one-out (LOO). See the text for definitions of these metrics. For each query allele, we measured each model's predictive performance (accounting for both sensitivity and specificity) by calculating an AUC value. The higher the AUC value the better the predictive performance. The plot shows the average difference between the AUC values for alleles belonging to the same locus obtained using our model vs. the corresponding values obtained using NetMHCIIpan, when similarity is defined based on either (A) the NN or (B) the LOO metric. Error bars denote standard deviations. Strikingly, our model performs better than NetMHCIIpan when predicting peptide binding to HLA-DQ using the NN metric (p-value $= 0.015$). For all other cases, both models have equivalent performance.
	}
	\label{fig5}
\end{figure}



We Compared how well our model predicts the MHC-II allele binding preferences of a novel peptide dataset vs. how well the state-of-the-art NetMHCIIpan method performs the same task. In this comparison we applied both our model and   NetMHCIIpan to predict binding preferences for peptides known to either bind or not bind a reference allele after training both models using peptide-binding data for a second allele. For a given MHC-II locus, the second allele was the one that was most similar to the reference allele. Similarity was quantified based on either  a leave-one-out approach or a nearest-neighbour approach (see Section\ref{nn}).
When using the nearest-neighbour approach, we found that our model performs significantly better than NetMHCIIpan in predicting peptide binding preferences for HLA-DQ alleles (P-value $= 0.015$; Figure \ref{fig5}, panel A).
 Furthermore, at   the $95\%$  confidence level, for all other cases, we found no significant difference between the performances of the two models (Figure \ref{fig5}).

These results are reassuring and indicate that our inverse-physics approach constitutes a promising  complement to the widely used pattern-based approach to peptide-MHC-II binding predictions. 
The outstanding predictive accuracy of the NetMHCIIpan is not the result of its theoretical basis. Rather it derives from the use of  sophisticated ensembles of neural networks, which are very powerful. However, our method has a distinguishing advantage over all the advanced machine learning models in that it is more physically meaningful.  It is worth noting that our prediction results of peptide-MHC-II interaction were based on \textit{in-silico} analysis of real data. Additional, in-vivo and in-vitro investigations are needed  to further validate the   reported predictive performance.

\bibliography{Trans_allele_model}

\end{document}


{\centering
		
		{\bfseries\Large Trans-Allelic Model for Prediction of Peptide:MHC-II Interactions \\ Supplementary Materials\bigskip}
		
		A. M. Degoot\textsuperscript{1,2,3} , Faraimunashe Chirove\textsuperscript{2} , and Wilfred Ndifon\textsuperscript{1} \\
		{\itshape
			\textsuperscript{1} African Institute of Mathematical Sciences-(AIMS), 6 Melrose, Muizenberg,  Cape Town, South Africa. \\
			\textsuperscript{2}School of Mathematics, Statistics and Computer Science,
			University of KwaZulu-Natal, Scottsville ave, Pietermaritzburg, South Africa \\
			\textsuperscript{3}DST-NRF Centre of Excellence in Mathematical and Statistical Sciences (CoE-MaSS), Wits 2050,  Gauteng,  South Africa  and \\
			\normalfont \today 
			
		}
	}


\section{Summary of the Optimization Algorithm}
Let $x_k$ be the  input vector that associated with a given pair of peptide $\mathsf{P^{(k)}}$ and MHC-II molecule $\mathsf{M^{(T(k))}}$, which can be extracted from 3-D structural information over all possible registers, and let $y_k \in \{0,1\}$ be the binary experimental value. For sake of an easy notation, let $\theta$ denotes for the model parameter (instead of $\Delta$ in the main text) representing the Hamiltonians.

The empirical loss function of \textbf{Eq\ref{e5} } in the main text will take the following form: 
\begin{equation}
\label{eq1}
\mathsf{G}_\mathsf{k}(\theta) = \mathsf{y}_\mathsf{k}\log(\pi_\mathsf{k}(\theta)) + (1 -\mathsf{y}_\mathsf{k})\log(1 - \pi_\mathsf{k}(\theta)),
\end{equation}
where 
$$ \pi_\mathsf{k}(\theta) = \frac{1}{1 + \exp(x_k^T\theta)}.$$
Equation \ref{eq1} could be simply written as follows:
\begin{equation}
	\label{eq2}
	\mathsf{G}_\mathsf{k}(\theta) =  (1-y_k)x_k^T\theta - \log(1 + \exp(x_k^T\theta)).
\end{equation} 

Then the learning function of  \textbf{Eq\ref{e4} } in the main text is:
\begin{equation}
\label{eq3}
\mathcal{L}( \mathsf{\theta}) = \underset{\{ \mathsf{\theta}\}}{\mathrm{argmin}}\left[-G(\theta) + \lambda||\theta||_1\right]= \sum_{k=1}^{N} \left((y_k -1)x_k^T\theta + \log(1 + \exp(x_k^T\theta)) \right) + \lambda\sum_{j=1}^{d}|\theta_j|,
\end{equation}
where $N$ is total number of the given data points and $d$ is the dimension of the parameter vector.

In order to get a closed formula for the update rule, we need to put the first sum of equation \ref{eq3}, $G(\theta)$,  in a quadratic form using Taylor expansion around a given point $\theta_0$.
\begin{equation}
\label{eq4}
G_Q(\theta) \approx G(\theta_0) + (\theta -\theta_0 )^T \frac{\partial G(\theta_0)}{\partial \theta} + \frac{1}{2}(\theta -\theta_0 )^T \frac{\partial^2 G(\theta_0)}{\partial \theta^2} (\theta - \theta_0) + C^2(\theta_0), 
\end{equation}
such that $C^2(\theta_0)$ is the truncated error of the second order in vicinity of the current point $\theta_0$. The first and second derivatives are:
$$ \frac{\partial G(\theta)}{\partial \theta} = \sum_{k=1}^{N} x_k^T\left(y_k -\bar{\pi}_k\right)$$ 
$$ \frac{\partial^2 G(\theta)}{\partial \theta^2} = \sum_{k=1}^{N} x_k^T\bar{\pi}_k(1-\bar{\pi}_k))x_k $$ 
where $\bar{\pi}_k = \pi_k(\theta_0)$ is the value of  probability  evaluated at point $\theta_0$.  Therefore, the quadratic approximation is:
\begin{equation}
\label{eq5}
G_Q(\theta) \approx  \sum_{k=1}^{N} \left[x_k^T(\theta -\theta_0)\bar{\pi}_k(1-\bar{\pi}_k))x_k^T(\theta -\theta_0)\right] + \sum_{k=1}^{N}\left[x_k^T(\theta -\theta_0)(y_k -\bar{\pi}_k\right)] + G(\theta_0) + C^2(\theta_0).
\end{equation}
Let $w_k = \bar{\pi}_k(1-\bar{\pi}_k)$, which is called re-weighting term, and re-write the above equation as follows:
\begin{equation}
	\label{eq6}
	G_Q(\theta) \approx  \frac{1}{2}\sum_{k=1}^{N} w_k\left[\left(x_k^T(\theta -\theta_0)\right)^2 + 2\frac{x_k^T(\theta -\theta_0)(y_k \bar{\pi}_k)}{w_k}\right] + G(\theta_0) + C^2(\theta_0).
\end{equation}
And even more simplified
\begin{equation}
\label{eq7}
G_Q(\theta) \approx \frac{1}{2}\sum_{k=1}^{N} w_k\left[ x_k^T(\theta -\theta_0) + \frac{y_k - \bar{\pi}_k}{w_k}\right]^2 + C(\theta_0),
\end{equation}
where $C(\theta_0)$ is an augmented constant, i.e. $C(\theta_0) = -\frac{1}{2}\sum_{k=1}^{N}\frac{y_k -\bar{\pi}_k}{w_k} +G(\theta_0) + C^2(\theta_0)$. Define the constant $z_k  = x_k^T\theta_0 + \frac{\bar{pi}_k -y_k}{w_k} $, and hence 
\begin{equation}
	\label{eq8}
	G_Q(\theta) \approx  \frac{1}{2}\sum_{k=1}^{N} w_k\left[z_k - x_k^T\theta \right]^2 + C(\theta_0).
\end{equation} 
By plugging equation \ref{eq8} into equation \ref{eq3} we get the a quadratic form of the  optimization function 
\begin{equation}
	\label{eq9} G_Q(\theta) \approx  \frac{1}{2}\sum_{k=1}^{N} w_k\left[z_k - x_k^T\theta \right]^2 + \lambda\sum_{j=1}^{d}|\theta_j|+ C(\theta_0).
\end{equation}
The above equation is convex but not differentiable; due to the  $L_1$ component, and we solved via iterative cyclic coordinate descent algorithm. 
$$\frac{\partial G_Q(\theta)}{\partial \theta_j} = 0 \Rightarrow -\sum_{k=1}^{N}\left[z_k -x_k^T \theta\right]x_{kj} + \lambda \frac{\partial|\theta_j|}{\partial \theta} = 0$$

$$ \Rightarrow \sum_{k=1}^{N}w_k(x_{kj})^2 - \sum_{k=1}^{N} w_k \left[z_k - \sum_{r=1;r\neq j}^{d}x_{kr}\theta_r \right]x_{kj} +\lambda \frac{\partial|\theta_j|}{\partial \theta} = 0.$$
For simplicity let $a_j = \sum_{k=1}^{N}w_k(x_{kj})^2$ and $b_j = \sum_{k=1}^{N} w_k \left[z_k - \sum_{r=1;r\neq j}^{d}x_{kr}\theta_r \right]x_{kj}$. Then 

$$ a_j \theta_j -b_j +\lambda \frac{\partial|\theta_j|}{\partial \theta} = 0.$$
Using idea of sub-differential 
$$ \begin{cases}
a_j\theta_j -b_j +\lambda, \hspace{0.2cm} \text{if} \hspace{0.2cm} \theta_j > 1\\
a_j\theta_j -b_j - \lambda,\hspace{0.2cm}  \text{if} \hspace{0.2cm} \theta_j <-1\\
0, \text{\hspace{0.6cm }otherwise} 
\end{cases}$$
As $a_j$ is absolutely non-negative, therefore the sign of $\theta_j$ depends on $\backepsilon_j$ and $\lambda$. Then 
\begin{equation}
	\label{eq10}
	\theta_j = \begin{cases}
	\frac{b_j - \lambda}{a_j}, \text{if} bj >0 \hspace{0.2cm} \text{and} \hspace{0.2cm} \lambda < |bj|\\
	\frac{b_j + \lambda}{a_j}, \text{if} b_j <0 \hspace{0.2cm} \text{and} \hspace{0.2cm} \lambda <|b_j|\\
	0, \hspace{0.6cm}\text{if} \lambda  \leq |b_j|.
	\end{cases}
\end{equation}
Equation \ref{eq10} known as soft-threshold operator.

The complete procedures of algorithm are following:
\begin{enumerate}
	\item Initialize  values for $\lambda, \lambda_{\text{min}}, \epsilon, \text{and} \hspace{0.2cm }\theta_0.$ 
	\item For every $j$ in the parameter vector 
	\subitem(a) Use the current value of $\theta$ to compute $a_j$ and $b_j$ for the $j$th component.
	\subitem(b) Use $a_j$ and $b_j$ to calculate new value for $\theta_j$ in equation \ref{eq10}.
    \subitem(c) Repeat (a) and (b) until convergent.
    \item Decrease $\lambda$ by $\lambda = \epsilon\lambda_{\text{current}}$
    \item While $\lambda > \lambda_{\text{min}} $ repeat 
\end{enumerate}

\textbf{Remark}: The re-weighting term $w_k = \bar{\pi}_k(1 -\bar{\pi}_k)$ might go to zero and  leads to a divergent results. In such cases we used its upper bound, i.e $w_k = 0.25$ (see \cite{glm}). 
\section{List of Tables}

\begin{center}
	\begin{longtable}{|C{3cm}| C{3.5cm}| L{6cm}| }
		
		\caption{ \textbf{Peptide:MHC-II complex structures}} \label{tab}\\
		\hline 
		\textbf{PDB Index} & \textbf{Allele} & \textbf{Peptide sequence}\\
		\endfirsthead
		\hline
		\endhead
		
		\multicolumn{3}{c}%
		{{\bfseries \tablename\ \thetable{} -- continued from previous page}} \\
		\hline
		\endhead
		
		\multicolumn{3}{c}{{Continued on next page}} \\
		\endfoot
		\endlastfoot 
		\hline
       1AQD  & $ DRB1^*01:01 $ 	& VGSDWRFLRGYHQYA\\ \hline
       1PYW	& $ DRB1^*01:01	$ 	& XFVKQNAAALX \\ \hline
       1KLG	& $ DRB1^*01:01	$ 	& GELIGILNAAKVPAD \\ \hline
       1KLU	& $ DRB1^*01:01	$ 	& GELIGTLNAAKVPAD \\ \hline
       2FSE	& $ DRB1^*01:01	$ 	& AGFKGEQGPKGEPG \\ \hline
       1SJH	& $ DRB1^*01:01	$ 	& PEVIPMFSALSEG \\ \hline
       1SJE	& $ DRB1^*01:01	$ 	& PEVIPMFSALSEGATP \\ \hline
       1T5W	& $ DRB1^*01:01 $ 	& 	AAYSDQATPLLLSPR \\ \hline
       1T5X	& $ DRB1^*01:01	$ 	& AAYSDQATPLLLSPR \\ \hline
       2IAN	& $ DRB1^*01:01	$ 	& GELIGTLNAAKVPAD \\ \hline
       2IAM	& $ DRB1^*01:01	$ 	& GELIGILNAAKVPAD \\ \hline
       2IPK	& $ DRB1^*01:01	$ 	& XPKWVKQNTLKLAT \\ \hline
       1FYT	& $ DRB1^*01:01	$ 	& PKYVKQNTLKLAT \\ \hline
       1R5I	& $ DRB1^*01:01	$ 	& PKYVKQNTLKLAT \\ \hline
       1HXY	& $ DRB^*10101	$ 	& PKYVKQNTLKLAT \\ \hline
       1JWM	& $ DRB1^*01:01	$ 	& PKYVKQNTLKLAT \\ \hline
       1JWS	& $ DRB1^*01:01	$ 	& PKYVKQNTLKLAT \\ \hline
       1JWU	& $ DRB1^*01:01	$ 	& PKYVKQNTLKLAT \\ \hline
       1LO5	& $ DRB1^*01:01	$ 	& PKYVKQNTLKLAT \\ \hline
       2ICW	& $ DRB1^*01:01	$ 	& PKYVKQNTLKLAT \\ \hline
       2OJE	& $ DRB1^*01:01	$ 	& PKYVKQNTLKLAT \\ \hline
       2G9H	& $ DRB1^*01:01	$ 	& PKYVKQNTLKLAT \\ \hline
       1A6A	& $ DRB1^*03:01	$ 	& PVSKMRMATPLLMQA \\ \hline
       1J8H	& $ DRB1^*04:01	$ 	& PKYVKQNTLKLAT \\ \hline
       2SEB	& $ DRB1^*04:01	$ 	& AYMRADAAAGGA \\ \hline
       1BX2	& $ DRB1^*15:01	$ 	& ENPVVHFFKNIVTPR \\ \hline
       1YMM	& $ DRB1^*15:01	$ 	& ENPVVHFFKNIVTPRGGSGGGGG \\ \hline
       1FV1	& $ DRB5^*01:01	$ 	& NPVVHFFKNIVTPRTPPPSQ \\ \hline
       1H15	& $ DRB5^*01:01	$ 	& GGVYHFVKKHVHES \\ \hline
       1ZGL	& $ DRB5^*01:01	$ 	& VHFFKNIVTPRTPGG \\ \hline
       4E41	& $ DRB1^*01:01 $ 	& 	GELIGILNAAKVPAD \\ \hline
       1DLH	& $ DRB1^*01:01	$ 	& PKYVKQNTLKLAT \\ \hline
       1KG0	& $ DRB1^*01:01	$ 	& PKYVKQNTLKLAT \\ \hline
       3L6F	& $ DRB1^*01:01	$ 	& APPAYEKLSAEQSPP \\ \hline
       3PDO	& $ DRB1^*01:01	$ 	& KPVSKMRMATPLLMQALPM \\ \hline
       3PGD	& $ DRB1^*01:01	$ 	& KMRMATPLLMQALPM \\ \hline
       3S4S	& $ DRB1^*01:01	$ 	& PKYVKQNTLKLAT  \\ \hline
       3S5L	& $ DRB1^*01:01	$ 	& PKYVKQNTLKLAT  \\ \hline
       1HQR	& $ DRB5^*01:01	$ 	& VHFFKNIVTPRTP \\ \hline
       3LQZ	& $ DPB1^*02:01 $ 	& RKFHYLPFLPSTGGS\\ \hline
       1UVQ	& $ DQB1^*06:02 $ 	& MNLPSTKVSWAAVGGGGSLV\\ \hline
       1JK8	& $ DQB1^*03:02 $ 	&LVEALYLVCGERGG\\ \hline
       1S9V	& $ DQB1^*02:01 $ 	& LQPFPQPELPY\\ \hline
       
\caption{Presents peptides MHC-II structural complexes collected from from references \cite{pan3.1, tepitopepan, hundi}. The first column gives the index of the protein databank, the second column presents the allele name of  the beta chain, and the third column shows the peptides corresponding to the allele.}
\end{longtable}
\end{center}

\newpage
	\begin{center}
	\begin{longtable}{|L{2.5cm}| L{2cm}| L{13cm}| }
		
		\caption{ \textbf{MHC-II polymorphic residue groups for HLA-DR genes.}} \label{tab1}\\
		\hline 
		Binding Pocket & \# of groups & polymorphic residue groups\\
		\endfirsthead
		\hline
		\endhead
		
		\multicolumn{3}{c}%
		{{\bfseries \tablename\ \thetable{} -- continued from previous page}} \\
		\hline
		\endhead
		
		\multicolumn{3}{c}{{Continued on next page}} \\
		\endfoot
		\endlastfoot 
		\hline
		\multicolumn{3}{|l|}{{\textbf{HLA-DRB molecules}}} \\
		\hline
		$P_1$ &  4  & \{82N, 85A, 86V, 89F\},  \{82N, 85V, 86G, 89F\}, \{82N, 85V, 86V, 89F \}, \{82Y, 85G, 86E, 89T\} \\ \hline
		$P_2$ &  5 &\{77T, 78Y, 81H, 82N\}, \{77N, 78Y, 81H, 82N\}, \{77T, 78V, 81H, 82N\}, \{77Y, 78C, 81N, 82Y\}, \{77T, 78Y, 81Y, 82N\}    \\ \hline
		$P_3$ & 8  & \{ 74A, 78Y\}, \{74E, 78V\}, \{74E, 78Y\}, \{74L, 78Y\}, \{74Q, 78V\}, \{74Q, 78Y\}, \{74R, 78Y\}, \{74V, 78C\}  \\ \hline
		$P_4$ & 20 & \{ 11L, 13F, 14E, 26L, 28E, 70Q, 71R, 74A, 78Y\}, \{ 11S, 13S, 14E, 26Y, 28D, 70Q, 71K, 74R, 78Y\}, \{ 11S, 13S, 14E, 26F, 28E, 70Q, 71K, 74R, 78Y\}, \{ 11V, 13H, 14E, 26F, 28D, 70Q, 71K, 74A, 78Y\}, \{ 11V, 13H, 14E, 26F, 28D, 70Q, 71R, 74A, 78Y\}, \{ 11G, 13Y, 14K, 26F, 28E, 70D, 71R, 74Q, 78V\}, \{ 11S, 13G, 14E, 26F, 28D, 70D, 71R, 74L, 78Y\}, \{ 11S, 13G, 14E, 26F, 28D, 70R, 71R, 74V, 78C\}, \{ 11D, 13E, 14F, 26Y, 28H, 70R, 71R, 74E, 78V\}, \{ 11S, 13S, 14E, 26F, 28D, 70D, 71R, 74A, 78Y\}, \{ 11S, 13G, 14E, 26L, 28E, 70D, 71R, 74A, 78Y\}, \{ 11S, 13S, 14E, 26F, 28D, 70D, 71E, 74A, 78Y\}, \{ 11S, 13S, 14E, 26F, 28E, 70Q, 71R, 74A, 78Y\}, \{ 11S, 13G, 14E, 26F, 28D, 70R, 71R, 74E, 78Y\}, \{ 11S, 13S, 14E, 26F, 28E, 70D, 71R, 74L, 78Y\}, \{ 11P, 13R, 14E, 26F, 28D, 70Q, 71A, 74A, 78Y\}, \{ 11R, 13S, 14E, 26Y, 28D, 70Q, 71K, 74R, 78Y\}, \{ 11L, 13S, 14E, 26F, 28E, 70Q, 71K, 74Q, 78Y\}, \{ 11A, 13C, 14E, 26N, 28I, 70R, 71R, 74E, 78Y\}, \{ 11D, 13Y, 14E, 26F, 28H, 70D, 71R, 74A, 78Y\}  \\ \hline
		$P_5$ & 20 & \{11L, 13F, 28E, 30C, 70Q, 71R, 74A\}, \{11S, 13S, 28D, 30Y, 70Q, 71K, 74R\}, \{11S, 13S, 28E, 30Y, 70Q, 71K, 74R\}, \{11V, 13H, 28F, 30Y, 70Q, 71K, 74A\}, \{11V, 13H, 28D, 30Y, 70Q, 71R, 74A\}, \{11G, 13Y, 28E, 30L, 70D, 71R, 74Q\}, \{11S, 13G, 28D, 30Y, 70D, 71R, 74L\}, \{11S, 13G, 28D, 30Y, 70R, 71R, 74V\}, \{11D, 13F, 28H, 30G, 70R, 71R, 74E\}, \{11S, 13S, 28D, 30Y, 70D, 71R, 74A\}, \{11S, 13G, 28E, 30H, 70D, 71R, 74A\}, \{11S, 13S, 28D, 30Y, 70D, 71E, 74A\}, \{11S, 13S, 28E, 30Y, 70Q, 71R, 74A\}, \{11S, 13G, 28D, 30Y, 70R, 71R, 74E\}, \{11S, 13S, 28E, 30Y, 70D, 71R, 74L\}, \{11P, 13R, 28D, 30Y, 70Q, 71A, 74A\}, \{11R, 13S, 28D, 30Y, 70Q, 71K, 74R\}, \{11L, 13S, 28E, 30Y, 70Q, 71K, 74Q\}, \{11A, 13C, 28I, 30Y, 70R, 71R, 74E\}, \{11D, 13Y, 28H, 30D, 70D, 71R, 74A\}  \\ \hline
		$P_6$ &  20 &\{9W, 11L, 13F, 28E, 30C, 70Q, 71R, 74A \}, \{9E, 11S, 13S, 28D, 30Y, 70Q, 71K, 74R \}, \{9E, 11S, 13S, 28E, 30Y, 70Q, 71K, 74R \}, \{9E, 11V, 13H, 28D, 30Y, 70Q, 71K, 74A \}, \{9E, 11V, 13H, 28D, 30Y, 70Q, 71R, 74A \}, \{9W, 11G, 13Y, 28E, 30L, 70D, 71R, 74Q \}, \{9E, 11S, 13G, 28D, 30Y, 70D, 71R, 74L \}, \{9E, 11S, 13G, 28D, 30Y, 70R, 71R, 74V \}, \{9K, 11D, 13F, 28H, 30G, 70R, 71R, 74E \}, \{9E, 11S, 13S, 28D, 30Y, 70D, 71R, 74A \}, \{9E, 11S, 13G, 28E, 30H, 70D, 71R, 74A \}, \{9E, 11S, 13S, 28D, 30Y, 70D, 71E, 74A \},\{9E, 11S, 13S, 28E, 30Y, 70Q, 71R, 74A \}, \{9E, 11S, 13G, 28D, 30Y, 70R, 71R, 74E \}, \{9E, 11S, 13S, 28E, 30Y, 70D, 71R, 74L \}, \{9W, 11P, 13R, 28D, 30Y, 70Q, 71A, 74A \}, \{9E, 11R, 13S, 28D, 30Y, 70Q, 71K, 74R \}, \{9E, 11L, 13S, 28E, 30Y, 70Q, 71K, 74Q \}, \{9E, 11A, 13C, 28I, 30Y, 70R, 71R, 74E \}, \{9Q, 11D, 13Y, 28H, 30D, 70D, 71R, 74A \}  \\ \hline
		$P_7$ & 21 & \{11L, 28E, 30C, 47Y, 61W, 67L, 70Q, 71R\}, \{11S, 28D, 30Y, 47F, 61W, 67L, 70Q, 71K\}, \{11S, 28E, 30Y, 47Y, 61W, 67L, 70Q, 71K\}, \{11V, 28D, 30Y, 47Y, 61W, 67L, 70Q, 71K\}, \{11V, 28D, 30Y, 47Y, 61W, 67L, 70Q, 71R\}, \{11G, 28E, 30L, 47Y, 61W, 67I, 70D, 71R\}, \{11S, 28D, 30Y, 47Y, 61W, 67F, 70D, 71R\}, \{11S, 28D, 30Y, 47Y, 61W, 67L, 70D, 71R\}, \{11S, 28D, 30Y, 47Y, 61W, 67L, 70R, 71R\}, \{11D, 28H, 30G, 47Y, 61W, 67F, 70R, 71R\}, \{11S, 28D, 30Y, 47F, 61W, 67F, 7D0, 71R\}, \{11S, 28E, 30H, 47F, 61W, 67I, 70D, 71R\}, \{11S, 28E, 30H, 47F, 61W, 67F, 70D, 71R\}, \{11S, 28D, 30Y, 47F, 61W, 67I, 70D, 71E\}, \{11S, 28E, 30Y, 47Y, 61W, 67L, 70Q, 71R\}, \{11S, 28E, 30Y, 47Y, 61W, 67L, 70D, 71R\}, \{11P, 28D, 30Y, 47F, 61W, 67I, 70Q, 71A\}, \{11R, 28D, 30Y, 47Y, 61W, 67L, 70Q, 71K\}, \{11L, 28E, 30Y, 47Y, 61W, 67L, 70Q, 71K\}, \{11A, 28I, 30Y, 47Y, 61W, 67L, 70R, 71R\}, \{11D, 28Y, 30D, 47Y, 61W, 67F, 70D, 71R\}  \\ \hline
		$P_8$ & 3 & \{60H, 61W\}, \{60S, 61W\}, \{60Y, 61W\}    \\ \hline
		$P_9$ &  14 & \{9W, 30C, 37S, 38V, 57D, 60Y, 61W\}, \{9E, 30N, 37Y, 38V, 57D, 60Y, 61W\}, \{9E, 30Y, 37Y, 38V, 57D, 60Y, 61W\}, \{9E, 30Y, 37Y, 38V, 57S, 60Y, 61W\}, \{9W, 30L, 37F, 38V, 57V, 60S, 61W\}, \{9E, 30Y, 37Y, 38V, 57I, 60Y, 61W\},   \{9K, 30N, 37g, 38V, 57V, 60S, 61W\}, \{9E, 30H, 37L, 38L, 57V, 60S, 61W\}, \{9E, 30Y, 37F, 38V, 57A, 60H, 61W\}, \{9W, 30Y, 37S, 38V, 57D, 60Y, 61W\}, \{9E, 30Y, 37F, 38L, 57V, 60S, 61W\},  \{9E, 30Y, 37F, 38V, 57V, 60S, 61W\}, \{9E, 30Y, 37Y, 38A, 57D, 60Y, 61W\},  \{9Q, 30D, 37D, 38L, 57D, 60Y, 61W\}          \\ \hline 
		\multicolumn{3}{|l|}{{\textbf{HLA-DP molecules}}} \\
		\hline
		$P_1$ & 2& \{86G, 89M\}, \{86D, 89V\} \\ 
		\hline
		$P_2$ & 2 & \{78M\}, \{78V\} \\
		\hline  
		$P_3$ & 2 & \{78M\}, \{78V\} \\
		\hline
		$P_4$ & 4 & \{71E\}, \{71K\}, \{71V\}, \{78V\} \\
		\hline
		$P_5$ & 2 & \{71E\}, \{71K\} \\
		\hline
		$P_6$ & 4 & \{9F\}, \{9Y\},\{71E\}, \{71K\} \\
		\hline
		$P_7$ & 2 & \{71E\}, \{71K\} \\
		\hline
		$P_8$ & - & \textbf{NA} \\
		\hline
		$P_9$ & 7 & \{9F \}, \{37F \}, \{38V\}, \{57D\}, \{9Y, 37Y\}, \{37L, 57E\}, \{38A, 57A \} \\
		\hline
		\multicolumn{3}{|l|}{{\textbf{HLA-DQ molecules}}} \\
		\hline
		$P_1$ & 2& \{86T, 89Q\}, \{86G, 89Q\} \\ 
		\hline
		$P_2$ & 1 &  \{78H\} \\
		\hline  
		$P_3$ & 1 &  \{78H\} \\
		\hline
		$P_4$ & 3 & \{71A, 78H\}, \{71E, 78H\}, \{71S, 78H\} \\
		\hline
		$P_5$ & 3 & \{71A\}, \{71E\}, \{71S\} \\
		\hline
		$P_6$ & 3 & \{9K, 71A\}, \{9K, 71E\}, \{9K, 71S\} \\
		\hline
		$P_7$ & 3 & \{71A\}, \{71E\}, \{71S\} \\
		\hline
		$P_8$ & - & \textbf{NA} \\
		\hline
		$P_9$ & 1 & \{9K, 37F, 38D, 57Y \}\\
		\hline
		\caption{Shows the polymorphic residue groups for  24 HLA-DRB, 5 HLA-DP and 6 HLA-DQ molecules for each of 9 binding pockets. The first column gives the nine binding pockets, the second column gives the number of polymorphic residue groups per each binding pocket, and the last column gives the polymorphic groups shown  by both the residue position number and amino acid type.}
	\end{longtable}
	
\end{center}
\begin{center}
	\begin{longtable}{|L{5cm}|C{5cm} |C{5cm} |}
		\caption{\textbf{Results of five-fold cross results for intra-allele vs trans-allele in terms of AUC values}}	\label{tabfold}\\
		\hline 
		\endfirsthead
		\endhead
		
		\multicolumn{3}{c}%
		{{\bfseries \tablename\ \thetable{} -- continued from previous page}} \\
		\endhead
		\hline
		\multicolumn{3}{c}{{Continued on next page}} \\
		\endfoot
		\endlastfoot
		\hline
		\textbf{Allele Name} &\multicolumn{2}{|c|}{{\textbf{AUC} }} \\\hline
		& \textbf{Intra-Allele training} & \textbf{Trans-Alleles training} \\ \hline
		\multicolumn{3}{|c|}{\textbf{HLA-DP molecules}} \\\hline
		$DPA1^*01:03-DPB1^*02:01$ &  $0.933$ & $0.933$ \\\hline
		$DPA1^*01:03-DPB1^*04:01$ &  $0.935$ & $0.939$ \\\hline
		$DPA1^*02:01-DPB1^*01:01$ &  $0.921$ & $0.922$ \\\hline
		$DPA1^*02:01-DPB1^*05:01$ &  $0.925$ & $0.926$ \\\hline
		$DPA1^*03:01-DPB1^*04:02$ &  $0.928$ & $0.927$ \\\hline
		\textbf{Overall}          &  $\mathbf{0.928}$       & $\mathbf{0.929}$\\\hline
		\textbf{p-value}          &   $\mathbf{0.594}$      &     $\mathbf{0.031}$   \\\hline
		\multicolumn{3}{|c|}{\textbf{HLA-DQ molecules}} \\\hline
		$DQA1^*01:01-DQB1^*05:01$ &  $0.864$ & $0.833$ \\\hline
		$DQA1^*01:02-DQB1^*06:02$ &  $0.830$ & $0.815 $ \\\hline
		$DQA1^*03:01-DQB1^*03:02$ &  $0.791$ & $0.728 $ \\\hline
		$DQA1^*04:01-DQB1^*04:02$ &  $0.883$ & $0.873 $ \\\hline
		$DQA1^*05:01-DQB1^*02:01$ &  $0.887$ & $0.871 $ \\\hline
		$DQA1^*05:01-DQB1^*03:01$ &  $0.884$ & $0.799 $ \\\hline
		\textbf{Average}          &   $\mathbf{0.857}$      & $\mathbf{0.820}$ \\\hline
		\textbf{p-value}          &   $\mathbf{0.4219}$      &   $\mathbf{0.047}$      \\\hline
		\multicolumn{3}{|c|}{\textbf{HLA-DRB molecules}} \\\hline
		$DRB1^*01:01$ &  $0.785$ & $0.786$ \\\hline
		$DRB1^*03:01$ &  $0.747$ & $0.725$ \\\hline
		$DRB1^*03:02$ &  $0.554$  & $0.657$ \\\hline
		$DRB1^*04:01$ &  $0.774$ & $0.756$ \\\hline
		$DRB1^*04:04$ &  $0.712$  & $0.744$ \\\hline
		$DRB1^*04:05$ &  $0.781$ & $0.794$ \\\hline
		$DRB1^*07:01$ &  $0.809$ & $0.825$ \\\hline
		$DRB1^*08:02$ &  $0.725$ & $0.716$ \\\hline
		$DRB1^*08:06$ &  $0.852$  & $0.880$ \\\hline
		$DRB1^*08:13$ &  $0.821$ & $0.837$ \\\hline
		$DRB1^*08:19$ &  $0.798$  & $0.790$ \\\hline
		$DRB1^*09:01$ &  $0.753$ & $0.757$ \\\hline
		$DRB1^*11:01$ &  $0.804$ & $0.833$ \\\hline
		$DRB1^*12:01$ &  $0.818$  & $0.824$ \\\hline
		$DRB1^*12:02$ &  $0.779$  & $0.848$ \\\hline
		$DRB1^*13:02$ &  $0.761$ & $0.718$ \\\hline
		$DRB1^*14:02$ &  $0.792$  & $0.818$ \\\hline
		$DRB1^*14:04$ &  $0.674$  & $0.674$ \\\hline
		$DRB1^*14:12$ &  $0.884$  & $0.878$ \\\hline
		$DRB1^*15:01$ &  $0.783$ & $0.784$ \\\hline
		$DRB3^*01:01$ &  $0.710$ & $0.674$ \\\hline
		$DRB3^*03:01$ &  $0.769$  & $0.765$ \\\hline
		$DRB4^*01:01$ &  $0.818$ & $0.802$ \\\hline
		$DRB5^*01:01$ &  $0.793$ & $0.820$ \\\hline
		\textbf{Average}&  $\mathbf{0.771}$ & $\mathbf{0.780}$         \\\hline
		\textbf{p-value}&       $\mathbf{0.169} $ & $\mathbf{0.329}$         \\\hline
		\caption{The first column gives the allele name and other two columns provide performance measurements of five-fold cross validation in term area under the ROC curve of-(AUC) \cite{auc} values for intra-allele, by applying our previous method \cite{ours}, and trans-allelic versions, respectively. Average and p-value (using Wilcoxon signed rank test \cite{daag}) statics were also provided for all three MHC-II allotypes.}
	\end{longtable}
\end{center}
	\begin{center}
	
	\begin{longtable}{|L{2.5cm}| L{2.5cm}| L{1.5cm}| L{1.5cm}|L{2.5cm}|L{1.5cm}|L{1.5cm}|L{1.5cm}|}
		\caption{Comparison results between nearest neighbourhood, Hamming distance and LOO method }	\label{tab3}\\
		\hline 
		\endfirsthead
		\endhead
		
		\multicolumn{8}{c}%
		{{\bfseries \tablename\ \thetable{} -- continued from previous page}} \\
		\endhead
		
		\multicolumn{8}{c}{{Continued on next page}} \\
		\endfoot
		\endlastfoot
		\hline
		\textbf{Query allele} &\textbf{ NN allele}    &\textbf{NN distance}&\textbf{NN AUC}& \textbf{H allele} &\textbf{H distanc}e& \textbf{H AUC}& \textbf{LOO AUC} \\ \hline
		\multicolumn{8}{|c|}{\textbf{HLA-DRB molecules}} \\
		\hline
		$DRB1^*01:01$& $DRB1^*15:01$&  0.056    &   0.703&$DRB1^*15:01$& 14 & 0.703 &0.735 \\ \hline  
		$DRB1^*03:01$& $DRB1^*03:02$&  0.013    &   0.553&$DRB1^*03:02$& 4  & 0.553 &0.680 \\ \hline 
		$DRB1^*03:02$& $DRB1^*03:01$&  0.013    &   0.618&$DRB1^*14:02$& 4  & 0.710 &0.625 \\ \hline
		$DRB1^*04:01$& $DRB1^*04:05$&  0.006    &   0.705&$DRB1^*04:05$& 2  & 0.705 &0.718 \\ \hline
		$DRB1^*04:04$& $DRB1^*04:01$&  0.008    &   0.724&$DRB1^*04:05$& 2  & 0.765 &0.752 \\ \hline
		$DRB1^*04:05$& $DRB1^*04:01$&  0.006    &   0.758&$DRB1^*04:04$& 2  & 0.722 &0.779 \\ \hline
		$DRB1^*07:01$& $DRB1^*09:01$&  0.058    &   0.748&$DRB1^*09:01$& 14 & 0.748 &0.789 \\ \hline
		$DRB1^*08:02$& $DRB1^*08:13$&  0.003    &   0.702&$DRB1^*08:13$& 1  & 0.702 &0.700 \\ \hline
		$DRB1^*08:06$& $DRB1^*08:02$&  0.009    &   0.739&$DRB1^*08:02$& 2  & 0.739 &0.830 \\ \hline
		$DRB1^*08:13$& $DRB1^*08:02$&  0.003    &   0.714&$DRB1^*08:02$& 1  & 0.714 &0.791 \\ \hline
		$DRB1^*08:19$& $DRB1^*08:13$&  0.017    &   0.819&$DRB1^*08:13$& 26 & 0.819 &0.790 \\ \hline
		$DRB1^*09:01$& $DRB1^*07:01$&  0.058    &   0.696&$DRB1^*07:01$& 14 & 0.696 &0.710 \\ \hline
		$DRB1^*11:01$& $DRB1^*13:02$&  0.002    &   0.649&$DRB1^*13:02$& 5  & 0.649 &0.804 \\ \hline
		$DRB1^*12:01$& $DRB1^*12:02$&  0.039    &   0.951&$DRB1^*12:02$& 1  & 0.951 &0.834 \\ \hline
		$DRB1^*12:02$& $DRB1^*12:01$&  0.039    &   0.954&$DRB1^*12:01$& 1  & 0.954 &0.848 \\ \hline
		$DRB1^*13:02$& $DRB1^*14:02$&  0.014    &   0.593&$DRB1^*11:01$& 5  & 0.633 &0.669 \\ \hline
		$DRB1^*14:02$& $DRB1^*03:02$&  0.013    &   0.675&$DRB1^*14:12$& 3  & 0.754 &0.794 \\ \hline
		$DRB1^*14:04$& $DRB1^*08:06$&  0.033    &   0.790&$DRB1^*14:12$& 8  & 0.696 &0.634 \\ \hline
		$DRB1^*14:12$& $DRB1^*14:02$&  0.014    &   0.777&$DRB1^*14:02$& 3  & 0.777 &0.854 \\ \hline
		$DRB1^*15:01$& $DRB1^*01:01$&  0.056    &   0.766&$DRB1^*01:01$& 14 & 0.766 &0.781 \\ \hline
		$DRB3^*01:01$& $DRB3^*03:01$&  0.033    &   0.579&$DRB3^*03:01$& 9  & 0.579 &0.590 \\ \hline
		$DRB3^*03:01$& $DRB3^*01:01$&  0.033    &   0.681&$DRB3^*01:01$& 9  & 0.681 &0.736 \\ \hline
		$DRB4^*01:01$& $DRB1^*04:04$&  0.120    &   0.698&$DRB1^*04:04$& 30 & 0.698 &0.740 \\ \hline
		$DRB5^*01:01$& $DRB1^*01:01$&  0.087    &   0.743&$DRB1^*01:01$& 23 & 0.743 &0.758 \\ \hline
		\textbf{Average} &              &           &   0.722&             &    & 0.727 &0.748 \\ \hline 
		\textbf{p-value} &              &           &   0.063&             &    & 0.060 &0.0612 \\ \hline 
		\multicolumn{8}{|c|}{\textbf{HLA-DP molecules}} \\ \hline
		$DPA1^*02:01-DPB1^*01:01$& $DPA1^*02:01-DPB1^*05:01$&  0.025    &   0.888&$DPA1^*02:01-DPB1^*05:01$& 8  & 0.888 &0.906 \\ \hline
		$DPA1^*01:03-DPB1^*02:01$& $DPA1^*03:01-DPB1^*04:02$&  0.005    &   0.929&$DPA1^*03:01-DPB1^*04:02$& 2  & 0.929 &0.937 \\ \hline
		$DPA1^*01:03-DPB1^*04:01$& $DPA1^*03:01-DPB1^*04:02$&  0.013    &   0.936&$DPA1^*03:01-DPB1^*01:01$& 4  & 0.941 &0.946 \\ \hline
		$DPA1^*03:01-DPB1^*04:02$& $DPA1^*01:03-DPB1^*02:01$&  0.005    &   0.910&$DPA1^*01:03-DPB1^*02:01$& 2  & 0.910 &0.926 \\ \hline
		$DPA1^*02:01-DPB1^*05:01$& $DPA1^*02:01-DPB1^*01:01$&  0.025    &   0.931&$DPA1^*02:01-DPB1^*01:01$& 8 & 0.931 &0.931 \\ \hline
		\textbf{Average} &              &           &  0.919&             &    & 0.920 &0.929 \\ \hline 
		\textbf{p-value} &              &           &   0.063&             &    &0.060 &0.0612 \\ \hline 
		\multicolumn{8}{|c|}{\textbf{HLA-DQ molecules}} \\ \hline
		$DQA1^*01:01-DQB1^*05:01$& $DQA1^*01:02-DQB1^*06:02$&  0.055    &   0.708&$DQA1^*01:02-DQB1^*06:02$& 17 & 0.708 &0.720 \\ \hline
		$DQA1^*01:02-DQB1^*06:02$& $DQA1^*01:01-DQB1^*05:01$&  0.055    &   0.515&$DQA1^*01:01-DQB1^*05:01$& 17 & 0.515 &0.684 \\ \hline
		$DQA1^*03:01-DQB1^*03:02$& $DQA1^*05:01-DQB1^*02:01$&  0.030    &   0.719&$DQA1^*05:01-DQB1^*03:01$& 7  & 0.581 &0.651 \\ \hline
		$DQA1^*04:01-DQB1^*04:02$& $DQA1^*03:01-DQB1^*03:02$&  0.044    &   0.763&$DQA1^*03:01-DQB1^*03:02$& 11 & 0.763 &0.728 \\ \hline
		$DQA1^*05:01-DQB1^*02:01$& $DQA1^*03:01-DQB1^*03:02$&  0.066    &   0.749&$DQA1^*03:01-DQB1^*03:02$& 19 & 0.749 &0.687 \\ \hline
		$DQA1^*05:01-DQB1^*03:01$& $DQA1^*03:01-DQB1^*03:02$&  0.030    &   0.653&$DQA1^*03:01-DQB1^*03:02$& 7  & 0.653 &0.619 \\ \hline
		\textbf{Average} &              &           &   0.685&             &    & 0.662 &0.682 \\ \hline 
		\textbf{p-value} &              &           &   0.0319&             &    & 0.030 &0.319 \\ \hline 
		\caption{ The first column gives the name of the query allele. \textbf{NN} allele and \textbf{H} allele  denote the nearest neighbour allele that is the most similar and has th shortest distance to the allele in query among  all other alleles in the training set, to nearest neighbour and Hamming distance approaches, respectively corresponding to the query allele. Likewise \textbf{NN} distance and \textbf{H} distance give the distance measurement between the two alleles calculated as described in Section \ref{nn} in the main text. The  columns  \textbf{NN AUC}, \textbf{H AUC} and \textbf{LOO AUC} show the prediction performance in terms of Area Under Curve-(AUC) \cite{auc} for nearest neighbour,  H distance and LOO approaches, respectively.  }
	\end{longtable}
\end{center}
\medskip 

\bibliography{Trans_allele_model}